# An Empirical Study of Touch-based Authentication Methods on Smartwatches

**Yue Zhao[*], Zhongtian Qiu[*], Yiqing Yang[*], Weiwei Li[*], Mingming Fan**
Department of Computer Science,
University of Toronto
{yuezhao, zqiu, yangyiq2, wwli194, mfan}@cs.toronto.edu

## ABSTRACT
The emergence of smartwatches poses new challenges to information security. Although there are mature touch-based authentication methods for smartphones, the effectiveness of using these methods on smartwatches is still unclear. We conducted a user study (n=16) to evaluate how authentication methods (*PIN* and *Pattern*), UIs (*Square* and *Circular*), and display sizes (*38mm* and *42mm*) affect authentication accuracy, speed, and security. Circular UIs are tailored to smartwatches with fewer UI elements. Results show that 1) *PIN* is more accurate and secure than *Pattern*; 2) *Pattern* is much faster than *PIN*; 3) *Square* UIs are more secure but less accurate than *Circular* UIs; 4) display size does not affect accuracy or speed, but security; 5) *Square PIN* is the most secure method of all. The study also reveals a security concern that participants' favorite method is not the best in any of the measures. We finally discuss implications for future touch-based smartwatch authentication design.

## Author Keywords
Authentication; security; smartwatch; empirical study.

## ACM Classification Keywords
H.5.2. Information Interfaces and Presentation: User Interfaces – Input devices and strategies, evaluation

## INTRODUCTION
Smartwatches are increasingly popular but also face unique challenges [10]. Although new methods are proposed to increase their I/O capability, their security is less addressed with most authentication methods being the same as those for smartphones (*e.g.,* [3, 5, 11, 14]). Recent studies further raised the security concern for smartwatches (*e.g.,* [12, 15]).

Compared with smartphones, ultra-small screen wearable devices have even smaller sized screens, various shapes, and are carried around in a different way from smart phones.

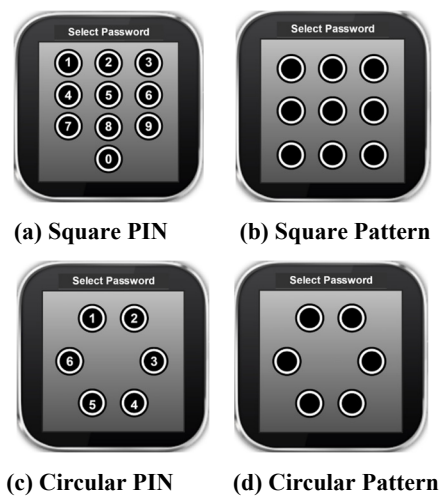

(a) Square PIN  (b) Square Pattern
(c) Circular PIN  (d) Circular Pattern

**Figure 1. Smartwatch touch-based authentication methods**

These might lead to different performance and preference of using commonly used touch-based authentication methods on smartwatches.

In this paper, we focus on touch-based authentication methods, as they are already familiar to mobile phone users, and do not require extra sensing devices (*e.g.,* high-quality camera for face recognition; bio-sensor for fingerprinting). We study how authentication methods (*PIN* and *Pattern*), user interfaces (*Square* and *Circular*) from the state-of-the-art, and display sizes (38mm & 42mm) (see Figure 1) affect the authentication accuracy, speed and security. A *PIN* method requires a 4-digit *PIN* password and a *Pattern* method requires a drawn pattern connecting dots on the interface. We use "*pattern with trace*" in this research, which is the most common *pattern* method on smartphones. For *Square* interfaces, a common 10-dot grid is used for *PIN* and a 9-dot grid is used for *Pattern*. *Circular* interfaces are based on a 6-dot circle, which are inspired by recently emerged smartphone authentication methods (*e.g.,* [11, 14]). *Circular* UIs may be a better fit for smaller sized smartwatches. We control all buttons to be the same size in all designs. In *Circular* layouts, we ensure all buttons are horizontally and vertically symmetric and evenly distributed.

Our study results contribute to the understanding of how touch-based authentication methods, UIs and display sizes affect the authentication accuracy, speed, security on smartwatches, and users' subjective preferences.

| Size | Authentication Scheme |
|---|---|
| 38mm | Circular PIN |
|  | Square PIN |
|  | Circular Pattern |
|  | Square Pattern |
| 42mm | Circular PIN |
|  | Square PIN |
|  | Circular Pattern |
|  | Square Pattern |

**Table 1. Eight authentication methods studied in this paper.**

## EXPERIMENT

We designed a 2*2*2 within-subject study to investigate the effects of three independent variables (IV) on three dependent variables. The three IVs are: the *authentication method* with two levels (*Pattern* and *PIN*); the *UI* with two levels (*circular* and *square*); *the* display *size* with two levels (*42mm* and *38mm)*, which are popular smartwatch sizes. Table 1 shows all the eight testing conditions. The three dependent variables are *input speed*, *input accuracy*, and *security*. To measure security, we conducted a shoulder surfing test, which is one of the most cited dangers for smartphone authentication systems [5]. We also measured *users' subjective preferences* via post-test questionnaires.

To simulate the smartwatch, we built an iOS application, which renders a smartwatch interface at the center of an iPhone 5S (see Figure 2(a)) with 2 different diameter sizes: 38mm (340*272 pixels) and 42mm (390*312 pixels). We stuck a wristband to the back of the smartphone so that participants could adjust it to wear the device firmly on their wrists (see Figure 2(b)). The weight of the simulator (112g) is comparative to many commercial smartwatches (e.g., Apple Watches:56g-125g). Similar to [6][7], we chose this approach to standardize the computing resource to eliminate any bias caused by the hardware and operating system.

### Tasks and Procedures

We recruited 16 volunteers (7 females), who were graduate students and professionals aged between 21 and 42 (*M=24, SD=7*) and all right-handed. Twelve used *PIN* and four used *Pattern* as passwords for their personal smartphones; the length of passwords that they used for their smartphones ranged from 4 to 7 (8 participants used 4-digit passwords; 6 used 6-digit ones; 2 used 5-digit ones; 2 used 7-digit ones). Only three of them used smartwatches before.

We conducted the experiment in a quiet office with each participant seated at a desk. To avoid the effect of fatigue, we limited the length of the experiment to be about 45 minutes in total. This dictates us to set the number of trials to be 15 for each of the 8 combinations (2 authentication methods * 2 UIs * 2 display sizes). We counterbalanced the authentication methods administered to participants using 8 * 8 Latin Square.

Each participant started with a training session and was asked to play around with all authentication methods until they felt comfortable. In the formal study, for each of the 8 authentication combinations, we asked participants to input

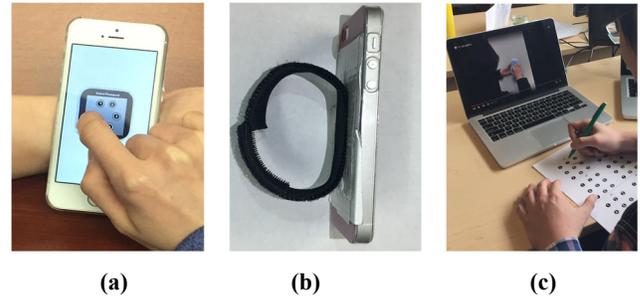

**Figure 2. (a) the simulated smartwatch. (b) a side view of the simulated smartwatch. (c) the shoulder surfing test.**

15 preset passwords/patterns, which were sequentially displayed on the simulated smartwatch. *PIN* passwords were randomly generated at the length of 4, the default length on both iOS and Android. *Patterns* were randomly generated at the length from 3 to 6 (the number of dots in a pattern) based on [2]. We did not ask participants to remember passwords, as it is infeasible to remember 15 * 8 combinations. To ensure that results of all trials are independent of each other and thus eliminate the effect of human corrections, our experimental software did not provide the performance feedback to participants immediately. However, an overall performance summary was shown to participants at the end.

After that, we conducted the shoulder-surfing test. We pre-recorded a video of a person entering 5 password sets using each of the 2*2*2 authentication methods (see Table 1). To record each video, we placed the camera 20 cm over the person's left shoulder for about half of the time and the person's right shoulder for the rest of the time from the same angle. During the test, we asked each participant to watch all pre-recorded videos on a 13' MacBook Pro and draw/write down the passwords/patterns that she/he saw in the videos (see Figure 2(c)). In the end, we asked participants to fill in a questionnaire to select the best authentication method in terms of efficiency, security, and overall preference.

### Measures and Data Collection

The authentication *accuracy* was measured by the average accuracy rate of the input passwords/patterns. The authentication *speed* was measured by the average completion time of the inputs. To guarantee the number of input is the same for every subject on each condition, both successful and unsuccessful inputs were considered. For each input, the time counter started when the user's finger touched on the screen and ended when the finger last lifted. *Security* was estimated by the rate of correctly replicated passwords/patterns in the shoulder surfing experiment.

For the accuracy and speed, the correctness of an authentication trial and time taken were recorded by the application. For security (the easiness of unauthorized replications), participants were asked to draw/write down the authentication patterns/pins observed from the pre-recorded simulated shoulder surfing videos. The number of correct replications was calculated by matching the attempts of participants and the actual patterns or pins.

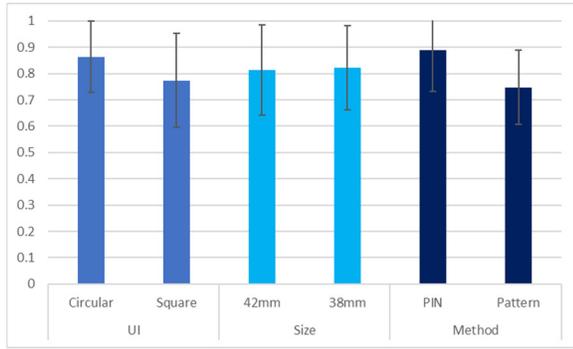

**Figure 3. Accuracies for all testing conditions**

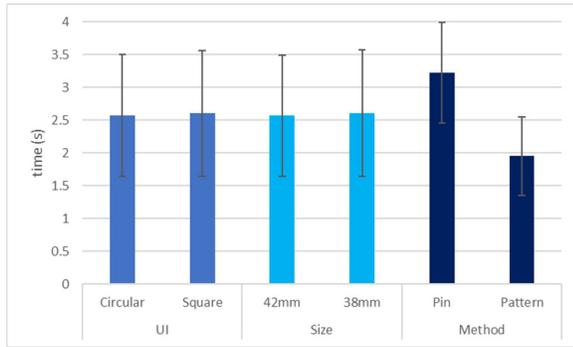

**Figure 4. Time spent for all testing conditions**

## RESULTS

The average authentication accuracy, speed and security are shown in Figure 3, Figure 4 and Figure 5 respectively.

We ran three-way repeated measure ANOVA on the accuracy, speed and security data respectively, and used the partial eta-squared ($\eta_p^2$) as a measure of the effect size. The results are shown in Table 2.

**Subjective Feedback:** We asked participants to select the best methods based on their perceived authentication speed, security, and overall preference. Most participants (77%) considered *Pattern* methods to be faster than *PIN* methods, with *Square Pattern* (39%) winning slightly more votes than *Circular Pattern* (38%). *Square UI* was considered more secure than *Circular UI*, whereas there was no bias between *Square Pattern* and *Square PIN*, both of which received 43% votes. Participants intuitively thought that more dots *(Square UIs)* could lead to better security. For the overall preference, *Pattern* methods were preferred by most participants compared to *PIN* methods. Specifically, *Square Pattern* was the most favorite method with 50% of participants' votes. 36% liked *Circular Pattern* most and the two *PIN* methods only got 7% votes each.

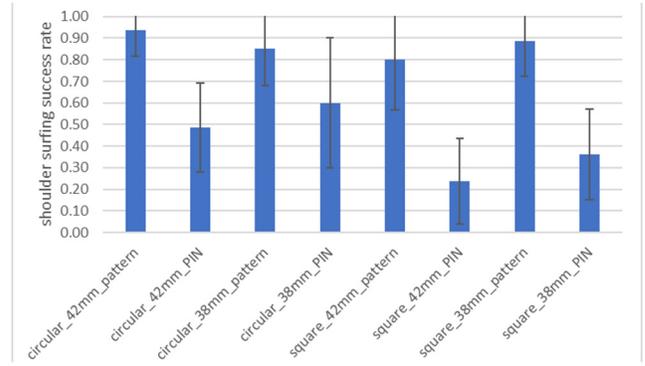

**Figure 5. Shoulder surfing attack success rates of all conditions**

## DISCUSSION

**Accuracy**: *PIN* methods are significantly more accurate than *Pattern* methods. This might be due to two factors:1) *Pattern* methods require longer finger motion paths on the screen, which increases the chance of failure; 2) the fat finger effect has a larger effect on *Pattern* methods. Twelve out of the sixteen participants (75%) complained that they had to move their fingers continuously on the small screen while drawing patterns, which increased the chance of visual occlusion. Results also show that *Circular* UI outperforms *Square* UI. This could be caused by the fewer buttons on the *Circular* UI than *Square* one (6 *vs.* 9).

**Authentication Speed**: *Pattern* methods have a significant advantage over *PIN* methods in speed. One possible reason is *Pattern* methods do not require repetitive touch and swipe actions on the screen. It is interesting to note that differences in *UI* or the *display size* do not significantly affect speed.

**Security**: All three independent variables and two interaction terms have a significant effect on security. Results reveal that *PIN* methods are more secure than *Pattern* methods (attack success rates: PIN (M:.42, SD:.26), Pattern (M:.87, SD:.18)). One explanation is that the trace of finger's motion is on the screen until the finger is lifted, which has less visual occlusion than repetitive touches in *PIN* methods and thus makes it easier for shoulder surfers to spy on. Another possible reason is memorizing a pattern is easier than remembering a password combination for most people [1, 4, 13]. Moreover, *Square* UI is more secure than *circular* UI (attack success rates: *Square* (*M:.57, SD:.34, Circular* (*M:.72, SD:.28*)). This is possibly due to more buttons on the Square UIs. Additionally, *Square PIN* is the most secure method of all.

**Table 2. Repeated measure ANOVA results on the effect of method, UI and size on the authentication accuracy, speed and security.**

|  | Accuracy | Speed | Security |
|---|---|---|---|
| Method (*PIN, Pattern*) | $F_{1, 15}$=52.71, $p < .05, \eta_p^2 = .78$ | $F_{1, 15}$=61.11, $p < .05, \eta_p^2 = .80$ | $F_{1, 15} = 125.35, p < .05, \eta_p^2 = .89$ |
| UI (*Square, Circular*) | $F_{1, 15}$=9.15, $p < .05, \eta_p^2 = .38$ | $F_{1, 15} = .09$, ns | $F_{1, 15} = 25.59, p < .05, \eta_p^2 = .63$ |
| Size (*38 mm, 42mm*) | $F_{1, 15}$=.35, ns | $F_{1, 15} = .35$, ns | $F_{1, 15} = 4.67, p < .05, \eta_p^2 = .24$ |
| Method * UI | $F_{1, 15} = .66$, ns | $F_{1, 15} = .21$, ns | $F_{1, 15} = .5.40, p < .05, \eta_p^2 = .27$ |
| Method * Size | $F_{1, 15} = .06$, ns | $F_{1, 15} = 4.39$, ns | $F_{1, 15} = 5.98, p < .05, \eta_p^2 = .29$ |
| UIs * Size | $F_{1, 15} = 2.90$, ns | $F_{1, 15} = .11$, ns | $F_{1, 15} = 2.48$, ns |
| Method * UIs * Size | $F_{1, 15} = .477$, ns | $F_{1, 15} = .08$, ns | $F_{1, 15} = 1.32$, ns |

**Questionnaire**: Results indicate most of the participants think *Pattern* methods require less authentication time, which is in line with our analysis. Participants thought *Square* methods are more secure than *Circular* methods, which is also consistent with our results. In terms of subjective preference, the most favorite method is *Square Pattern*. One possible reason is that people are more familiar with it since it is one of the dominant authentication methods on smartphones [3]. However, this raises a security concern because *Square Pattern* is not the best in any measures.

**The Size of Smartwatches**: Results show that the size of smartwatch (18.14% difference in the touching area) does not have significant effect on the authentication accuracy and speed. It contradicts with the intuition that larger watch face might lead to the increase in accuracy because of the slightly more space for fingers to touch, the decrease in the authentication speed because of the longer distance for fingers to travel. Additionally, the larger watch size did not lead to the decrease in the security (attack success rates *:42mm (M:.62, SD:.33), 38mm (M:.68, SD:.30))*. However, whether this will hold true for larger differences in the sizes of smartwatches needs further examination.

**Limitations**: We conducted the study on a simulated smartwatch. Although this design keeps external factors (*e.g.,* hardware, OS) constant, it might still deviate from a real smartwatch due to the differences in weight and convenience. Second, in shoulder surfing tests, we only considered a replication as a success if the fully correct password was given. However, hackers may get a partially correct password and use it to increase the possibility of brute-force attack. Thus, the degree of similarity between the entered password and the correct one may be considered. In addition to the entered password, the way how people enter them can also be considered, which was shown as a promising enhancement to password patterns to decrease the risk of side-channel attack (*e.g.,* [8, 9]). Third, our study evaluated *pattern with trace* since it is more popular than *pattern without a trace*. However, *pattern without a trace* may increase the difficult for shoulder surfing attack. Fourth, our study examined the common shapes for the circular and square UIs with the same sized buttons. However, the number of buttons in these UIs are different. UI designs with the same number of buttons for *Square* and *Circular* UIs correspondingly can be studied to better generalize the findings between *Square* and *Circular* UIs. Additionally, there are other uncommon shapes for organizing the buttons, such as hexagon, octagon, which may affect the touch-based authentication.

Our work is an initial exploration of factors that may affect the performance of touch-based authentication methods on smartwatches. Besides the limitations mentioned above, many research questions remain open. For example, users' prior experience of using PIN/Pattern methods, their physical state (*e.g.,* sitting, standing and walking) can also play a role in users' performance and preference of the authentication methods.

## CONCLUSION

We empirically studied how two watch sizes (*38mm* and *42mm*), two UIs (*Square* and *Circular*) and two methods (*PIN* and *Pattern*) affected accuracy, speed, and security of the touch-based smartwatch authentication. Results show that:1) *PIN* are more accurate and secure than *Pattern*; 2) *Pattern* are much faster than *PIN*; 3) *Square* UIs are more secure but less accurate than *Circular* UIs, and UIs do not affect speed significantly; 4) the display size does not affect accuracy or speed, but security; 6) *Square PIN* is the most secure method of all; 7) users' most preferred method is *Square Pattern,* which is not the best in any measure. As the first empirical validation work on touch-based authentication methods on smartwatches, our results provide insights for app developers in designing authentication methods by considering tradeoffs between accuracy, speed and security.

## REFERENCES

[1] Andriotis, P., et al. 2014. Complexity metrics and user strength perceptions of the pattern-lock graphical authentication method. *HAISPT'14*, 115–126.

[2] Andriotis, P., et al. 2013. A Pilot Study on the Security of Pattern Screen-lock Methods and Soft Side Channel Attacks. *WiSec'13*, 1–6.

[3] Arif, A. and Mazalek, A. 2014. Slide-to-Unlock Revisited: Two New User Authentication Techniques for Touchscreen-Based Smartphones. *MOBIQUITOUS'14*, 389–390.

[4] Chiasson, S., et al. 2009. Multiple password interference in text passwords and click-based graphical passwords. *CCS'09*, 500-511.

[5] Harbach, M., et al. 2014. It's a hard lock life: a field study of smartphone (un) locking behavior and risk perception. *SOUPS'14*, 213–230.

[6] Hsiu, M.-C., et al. 2016. ForceBoard: using force as input technique on size-limited soft keyboard. *MobileHCI'16*, 599–604.

[7] Leiva, L.A., et al. 2015. Text Entry on Tiny QWERTY Soft Keyboards. *CHI'15*, 669–678.

[8] De Luca, A., et al. 2012. Touch me once and i know it's you!: implicit authentication based on touch screen patterns. *CHI'12*, 987-996.

[9] Maiti, A., et al. 2015. (Smart)watch your taps: side-channel keystroke inference attacks using smartwatches. *ISWC'15*, 27–30.

[10] Rawassizadeh, R., et al. 2014. Wearables: has the age of smartwatches finally arrived? *Communications of the ACM*. 58, 1, 45–47.

[11] Shin, K.I., et al. 2012. Design and implementation of improved authentication system for Android smartphone users. *WAINA'12*, 704-707.

[12] Shukla, D., et al. 2014. Beware, Your Hands Reveal Your Secrets! *CCS'14*, 904–917.

[13] Suo, X., et al. 2005. Graphical passwords: a survey. *ACSAC'05*, 10–19.

[14] Truong, K.N., et al.. 2014. Slide to X: unlocking the potential of smartphone unlocking. *CHI'14*, 3635–3644.

[15] Wang, H., et al. 2015. MoLe: Motion Leaks through Smartwatch Sensors. *MobiCom'15*, 155–166.